# Phase tuning in Michelson-Morley Experiments performed in vacuum, assuming length contraction

Joseph Levy
4 Square Anatole France, 91250 St Germain lès Corbeil, France
Levy.joseph@orange.fr

In agreement with Michelson-Morley experiments performed in vacuum, we show that, assuming the existence of a fundamental aether frame and of a length contraction affecting the material bodies in the direction of the Earth absolute velocity, the light signals, travelling along the arms of the interferometer arrive in phase whatever their orientation, a result which responds to an objection opposed to the non-entrained aether theory. This result constitutes a strong argument in support of length contraction and of the existence of a model of aether non-entrained by the motion of celestial bodies.

**I. Introduction**

   It is well established today that no fringe shift occurs when Michelson and Morley experiments are performed in vacuum [1-5]. The explanation of this result, based on Lorentz and Fitzgerald assumptions, attributes the effect to a length contraction affecting the material bodies in the direction of their absolute motion. The said hypothesis is supported by the fact that it perfectly predicts that the two–way transit time of light along a rod surrounded by the vacuum, is independent of the angle separating the rod from any fixed axis [6-9].
   However the objection which was done to Lorentz's hypothesis, is that the light signals that propagate along the two arms of the interferometer have little chance to be in phase when they reach the detector, whatever the orientation of the arms.
   We propose in this text to check the exactness (or not) of this claim.
   Given the equality of the two-way transit time of light mentioned above, it is clear that if the number of oscillations in the two arms is equal, the waves will be in phase when they reach the detector. We will verify whether, assuming length contraction in the direction of the absolute motion of the Earth frame, the number of oscillations in the round trip of a light signal along a rod is (or not) dependent on its orientation , and the implications of this fact on Michelson-Morley experiment.[1]

   Consider to this end a rod at rest on Earth surrounded by a vacuum, whose direction makes an angle $\theta$ relative to the Earth absolute velocity vector, and let a monochromatic electromagnetic wave travel back and forth along the rod. Due to length contraction in the direction of the Earth absolute velocity, the length of the rod varies as a function of the angle. The wavelength also varies as a result of a Doppler effect, and its value differs in the forward and in the backward direction. Denoting by $\lambda$ and $\lambda'$ the wavelengths of the signal in the Earth frame in the two opposite directions, the number of oscillations in a round trip will be:

$$N + N' = \frac{\ell}{\lambda} + \frac{\ell}{\lambda'}, \qquad (1)$$

where $\ell$ is the real length assumed by the rod in the direction $\theta$ (whose value will be determined in paragraph II.)

---

[1] The demonstration concerns all the orientations of the rod. It has been verified experimentally for the Michelson-Morley vacuum experiments where the arms lie in a plane parallel to the surface of the Earth and this whatever the moment of the day [1-3]. Since the orientation of the interferometer varies during the 24 hours of the day, one can conclude that the phase tuning is verified whatever the direction of the arms (This diurnal change of orientation occurs in the absence of gravity variation, allowing to test the effect of the aether and of length contraction only. See also the footnote 2 page 6).

On the other hand, during the forward transit along the rod, the wave covers also a path $\ell_0$ in the aether frame. Of course, the number of oscillations $N$ is the same in $\ell_0$ and in $\ell$. Denoting by $\lambda_0$ the wavelength in the aether frame, we have:

$$\frac{\ell}{\ell_0} = \frac{N\lambda}{N\lambda_0} = \frac{C't}{Ct}$$

so that

$$\lambda = \lambda_0 \frac{C'}{C} \qquad (2)$$

where $C$ is the speed of light in the aether frame, (which is independent of the direction of the light signal).
$C'$ is the real speed relative to the Earth frame of the light signal that travels from the origin until the end of the rod (forward direction), measured with clocks exactly synchronized and not slowed down by motion. ($C'$, which varies as a function of the angle, will be determined in section III).

The origins of the paths $\ell$ and $\ell_0$, coincide, when the light signal meets them, this is also the case when it meets the path's ends. Therefore the time needed by the light signal to cover these paths is the same. Clocks at rest with respect to one another and perfectly synchronous should thus display the same reading for both paths. Measured with such clocks not slowed down by motion, this reading is the real time $t$.

If the rod is aligned in the direction of the Earth absolute velocity, the real speed of light in the forward direction becomes $C'=C-v$, so that expression (2) reduces to:

$$\lambda = \lambda_0 (1 - v/C),$$

which is the expression of the classical Doppler effect in the forward direction.

During the backward transit along the rod, the wave covers the path $\ell'_0$ in the aether frame. The number of oscillations $N'$ is the same in $\ell'_0$ and in $\ell$, so that:

$$\frac{\ell}{\ell'_0} = \frac{N'\lambda'}{N'X} = \frac{C''t'}{Ct'}, \qquad (3)$$

where $C''$ is the real backward speed of light in $\ell$ relative to the Earth frame, and $t'$ is the real backward transit time of the light signal in $\ell$ and in $\ell'_0$.

$X$ designates the wavelength of the light signal in $\ell'_0$ after reflection.

For a given orientation the length of the rod does not vary, therefore, the period $\tau$ of the signal remains the same before and after reflection. We thus have from expression (2):

$$C'\tau = \lambda_0 \frac{C'}{C} \Rightarrow \tau = \frac{\lambda_0}{C}$$

and from expression (3):

$$\lambda' = C''\tau = \frac{C''}{C} X$$

which yields:

$$X = C\tau = \lambda_0.$$

Therefore:
$$\lambda' = \lambda_0 \frac{C''}{C}. \qquad (4)$$

In the direction opposite to the Earth absolute velocity, the speed of light becomes $C''=C+v$, so that expression (4) reduces to:
$$\lambda' = \lambda_0 (1 + v/C)$$
which is the expression of the classical Doppler effect in the direction opposite to the Earth absolute motion.

Taking account of (1), (2) and (4), we can see that:
$$N + N' = \frac{\ell C}{\lambda_0}(\frac{1}{C'} + \frac{1}{C''}) = \frac{\ell C}{\lambda_0}(\frac{C'+C''}{C'C''}). \qquad (5)$$

The complete determination of the number of oscillations, in the two-way transit of the light signal, needs therefore assessment of $\ell, C'$ and $C''$.

## II. Length of a rod whose projection in the *x-x'* direction is contracted

Consider two co-ordinate systems $S_0$ and $S_1$. $S_0$ is at rest in the aether frame, and $S_1$ moves uniformly at speed *v* along the *x*-axis of the co-ordinate system $S_0$. A long rod having its origin in O' is aligned in a direction which can be different from the *x,x'*-axis (Figure1).

Assuming that length contraction occurs according to Fitzgerald and Lorentz in the *x,x'*-direction of the absolute velocity vector, we will determine the length of the rod whose projection in the direction of the *x,x'*-axis is contracted.

(In accordance with the conditions expressed in the footnote (1) of the first page, we only consider the cases where no gravity variation occurs).

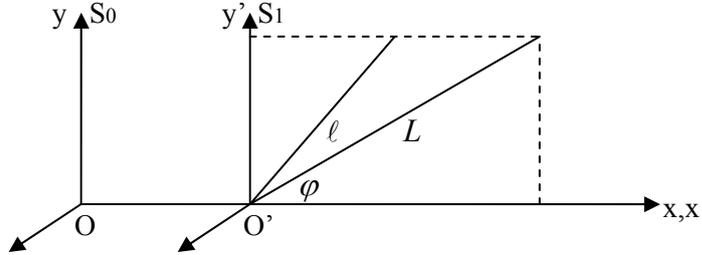

Figure 1. Along the *x, x'*-axis, the projection of the rod *L* contracts, along the y'-axis it is not modified.

*L* is the length assumed by the rod when it is at rest in the aether frame, and $\ell$ its real length in the moving co-ordinate system. Along the *x,x'*-axis, we have:
$$L\cos \varphi = \frac{\ell \cos \theta}{\sqrt{1 - v^2/C^2}},$$
and along the y'-axis:
$$L \sin \varphi = \ell \sin \theta,$$
where $\varphi$ is the angle separating *L* from the *x,x'*-axis, and $\theta$ the angle between the *x,x'*- axis and $\ell$.

**Important note**

In the figure, the issue is addressed in the *x,y* plane, but the same results would be obtained in the general case for any plane passing by the *x,x'*-axis.

From the Pythagorean law we have:

$$\left(\frac{\ell \cos\theta}{\sqrt{1 - v^2/C^2}}\right)^2 + (\ell \sin\theta)^2 = L^2.$$

Thus:

$$\ell = \frac{L(1 - v^2/C^2)^{1/2}}{(1 - v^2 \sin^2\theta/C^2)^{1/2}}. \tag{6}$$

It is important to realize that $\ell$ is the real length of the rod in $S_1$ but it is not the measured length in this co-ordinate system. Indeed, the standards used to measure the rod are also contracted in the same ratio, and, therefore, its *apparent* length in the co-ordinate system $S_1$ is found to be $L$.

Conversely, an observer in the system $S_0$ would have obtained an exact estimation of the length of the rod because the standards in $S_0$ are not contracted.

## III. Determination of C' and C" [6].

Let us consider the two co-ordinate systems, $S_0$ and $S_1$ mentioned above [6-9], and the rod $\ell = OA$ firmly fixed to the co-ordinate system $S_1$ (Figure 2). (We specify that the study we will deal with, relates to measurements that are all carried out in vacuum).

Let us place, at the two ends of the rod, two mirrors facing one another by their reflecting surfaces which are perpendicular to the axis of the rod. At the initial instant, the two systems $S_0$ and $S_1$ overlap. At this very instant a light signal is sent from the common origin and travels along the rod towards point A. When the signal reaches this point the rod has been transferred to a distance equal to $OO' = vt$ and is referred to as *O'A'* where *t* is the time needed by the signal to cover the distance *O'A'*.

(See figure 2).

After reflection the signal reverses its travel. (Note that the length of the moving rod is contracted according to formula (6)).

We remark that the path of the light signal along the rod is related to the speed $C'$ by the relation:

$$C' = \frac{O'A'}{t}.$$

As we saw, when the signal reaches point A', the system $S_1$ has moved away from $S_0$ a distance *OO'=vt*, so that:

$$v = \frac{OO'}{t}.$$

The same distance has been covered by point A which is transferred to A'.

Now, from the point of view of an observer which is supposed at rest in $S_0$, the signal goes from point O to point A' (see figure 2).

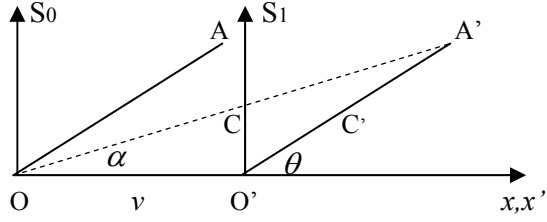

Figure 2. The speed of light is equal to C' from O' to A' and to C from O to A'.

$C$ being the speed of light in $S_0$, we have:

$$\frac{OA'}{t} = C,$$

and hence, the projection along the $x,x'$-axis of the speed of light $C'$ relative to the system $S_1$, will be equal to $(C\cos\alpha - v)$. So that:

$$C\cos\alpha - v = C'\cos\theta.$$

The three speeds, $C$, $C'$, and $v$ being proportional to the three lengths OA', OA and OO' with the same coefficient of proportionality, we have

$$C^2 = (C'\cos\theta + v)^2 + C'^2 \sin^2\theta.$$

Therefore:

$$C'^2 + 2vC'\cos\theta - (C^2 - v^2) = 0. \qquad (7)$$

(We must emphasize that equation (7) implies that the three speeds $C$, $C'$ and $v$ have been measured with the help of clocks which obviously are not slowed down by motion, and which display the same time $t$).

Resolving the second degree equation, yields:

$$C' = -v\cos\theta \pm \sqrt{C^2 - v^2\sin^2\theta}.$$

The condition $C' = C$ when $v = 0$ compels us to only retain the + sign so:

$$C' = -v\cos\theta + \sqrt{C^2 - v^2\sin^2\theta}. \qquad (8)$$

QED

Now, the return of light can be illustrated by the figure 3 below:

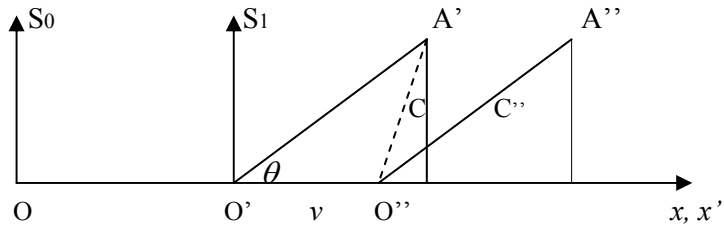

Figure 3. The speed of light is equal to C'' from A'' to O'' and to C from A' to O''. $\alpha'$ is the angle separating A'O'' from the $x,x'$-axis. (not indicated in the figure)

From the point of view of an observer attached to the system $S_1$, the light comes back to its initial position with the speed $C''$. Therefore we can write:

$$C'' = \frac{A''O''}{t'},$$

where $t'$ is the time of light transit from its final to its initial position.

For an observer which is supposed at rest relative to frame $S_0$, the light comes from A' to O'' with the speed $C$, so that:

$$C = \frac{A'O''}{t'}.$$

During the light transit, the system $S_1$ has moved from O' to O'' with the speed $v$ therefore:

$$v = \frac{O'O''}{t'}.$$

The projection of the speed of light relative to $S_1$ along the $x,x'$-axis will be:

$$C\cos\alpha' + v = C''\cos\theta$$

where $\alpha'$ is the angle separating A'O'' from the $x,x'$-axis.

We easily verify that:

$$(C''\cos\theta - v)^2 + (C''\sin\theta)^2 = C^2,$$

therefore,

$$C''^2 - 2vC''\cos\theta - (C^2 - v^2) = 0. \qquad (9)$$

(Expression (9) also implies that the three speeds, $C$, $C''$ and $v$, have been measured with the help of clocks not slowed down by motion, and which display the same time $t'$).

Resolving the second degree equation yields:

$$C'' = v\cos\theta + \sqrt{C^2 - v^2\sin^2\theta}. \qquad (10)$$

QED

Note that during a short time the Earth frame can be considered almost inertial. If this was not the case, we would be affected by the accelerations. So, in the study we deal with, the Earth frame can be identified to $S_1$.

### IV. Number of oscillations during a round trip of the light signal along the rod

From (5), (6), (8) and (10) we obtain:

$$N + N' = \frac{LC\sqrt{1 - v^2/C^2}}{\lambda_0 \sqrt{1 - v^2\sin^2\theta/C^2}} \frac{(C' + C'')}{C'C''}$$

$$= \frac{L\sqrt{1 - v^2/C^2}}{\lambda_0 \sqrt{1 - v^2\sin^2\theta/C^2}} x \frac{2C^2\sqrt{1 - v^2\sin^2\theta/C^2}}{C^2 - v^2}.$$

Finally:

$$N + N' = \frac{2L}{\lambda_0 \sqrt{1 - v^2/C^2}}.$$

In conclusion, we see that, assuming the existence of a fundamental aether frame and length contraction, the number of oscillations in the round trip of a light signal along a rod standing in the Earth frame which is surrounded by a vacuum, does not depend on the orientation of the rod[2]. So, the waves travelling in the transverse arm and the longitudinal arm, in the Michelson-Morley interferometer, are in phase when they reach the telescope detector, and this, whatever the direction of the arms, in agreement with experiment. This constitutes a weighty argument in support of length contraction and gives a response to those who assert that, with this assumption, the phase tuning does not occur for all orientations of the interferometer. Moreover it shows that the non-entrained aether theory rests on firmly established arguments.

---

[2] F.G. Pierce has found a fringe shift in a vertical orientation, but the experiment was not performed in vacuo [10]. One might also wonder whether the gravitational field could be responsible for a fringe shift? M.A.F. Rosa and W.A. Rodriguez Jr carried out a rigorous theoretical search of the expected result. No fringe shift was predicted by these authors [11].

**Acknowledgements**

Many thanks to Ian Montgomery and Guy Grantham, for drawing my attention to the subject and for their interest in the conclusions of this work.
Many thanks also to Franco Selleri, for the positive review he provided.